\documentstyle[11pt,aaspp4,flushrt,tighten]{article}
\singlespace

\def\nH{{\rm H}}
\def\nHI{{\rm H\,I}}

\def\HeII{\hbox{He~$\scriptstyle\rm II\ $}}

\def\kms{\,{\rm km\,s^{-1}}}
\def\kmsmpc{\,{\rm km\,s^{-1}\,Mpc^{-1}}}
\def\cmm{\,{\rm cm^{-2}}}
\def\uvunits{\,{\rm ergs\,cm^{-2}\,s^{-1}\,Hz^{-1}\,sr^{-1}}}
\def\xrbunits{\,{\rm keV\,cm^{-2}\,s^{-1}\,keV^{-1}\,sr^{-1}}}
\def\eden{\,{\rm eV\,cm^{-3}}}

\def\Lya{Ly$\alpha\ $}
\def\etal{{et al.\ }}
\def\epsi{{\epsilon}}
\def\spose#1{\hbox to 0pt{#1\hss}}
\def\lta{\mathrel{\spose{\lower 3pt\hbox{$\mathchar"218$}}
     \raise 2.0pt\hbox{$\mathchar"13C$}}}
\def\gta{\mathrel{\spose{\lower 3pt\hbox{$\mathchar"218$}}
     \raise 2.0pt\hbox{$\mathchar"13E$}}}
\begin{document}
\title{Compton Heating of the Intergalactic Medium by the Hard X--ray 
Background}
\author{Piero Madau\altaffilmark{1}}
\affil{Space Telescope Science Institute, 3700 San Martin Drive, Baltimore, 
MD 21218}
\and
\author{George Efstathiou}
\affil{Institute of Astronomy, Madingley Road, Cambridge CB3 0HA, UK}

\altaffiltext{1}{also at Institute of Astronomy, Madingley Road, Cambridge 
CB3 0HA, UK.}
 
\begin{abstract}
High--resolution hydrodynamics simulations of the \Lya forest in cold
dark matter dominated cosmologies appear to predict line--widths that
are substantially narrower than those observed. Here we point out that
Compton heating of the intergalactic gas by the hard X--ray background
(XRB), an effect neglected in all previous investigations, may help to
resolve this discrepancy. The rate of gain in thermal energy by
Compton scattering will dominate over the energy input from hydrogen
photoionization if the XRB energy density is $\sim 0.2\,x/\langle
\epsilon\rangle$ times higher than the energy density of the UV
background at a given epoch, where $x$ is the hydrogen neutral
fraction in units of $10^{-6}$ and $\langle \epsilon\rangle$ is the
mean X--ray photon energy in units of $m_ec^2$. The numerical
integration of the time--dependent rate equations shows that the
intergalactic medium approaches a temperature of about $1.5\times
10^4\,$ K at $z>3$ in popular models for the redshift evolution of the
extragalactic background radiation. The importance of Compton heating
can be tested experimentally by measuring the \Lya line--width
distribution as a function of redshift, thus the \Lya forest may
provide a useful probe of the evolution of the XRB at high redshifts.

\end{abstract}


\section{Introduction}

Neutral hydrogen in the intergalactic medium (IGM) produces a plethora
of \Lya absorption lines in the spectra of high redshift quasars.
Numerical N--body/hydrodynamics simulations of structure formation in
the IGM, within the framework of cold dark matter (CDM) dominated
cosmologies (Cen \etal 1994; Zhang, Anninos, \& Norman 1995;
Miralda--Escud\'e \etal 1996; Hernquist \etal 1996; Zhang \etal 1998;
Theuns \etal 1998a,b; Dav\'e \etal 1999; Bryan \etal 1999), have
recently provided a definite picture for the origin of the \Lya
forest, one of an interconnected network of sheets and filaments with
virialized systems (halos) located at their points of intersection. At $z>2$, 
the lowest column density absorbers ($N_{\rm H\,I}\sim 10^{12}\cmm$) arise in 
this context in the underdense ($\rho_b/\bar \rho_b<1$)
minivoids between the filaments, absorbers that give rise to moderate
column density lines ($N_{\rm H\,I}\lta 10^{14} \cmm$) correspond to
modestly overdense ($1<\rho_b/\bar \rho_b<5$) filaments, and high
column density absorbers ($N_{\rm H\,I}\gta 10^{15}\cmm$) arise in highly
overdense ($\rho_b/\bar \rho_b>10$) structures at the intersection of
filaments (Zhang \etal 1998). Most simulations to date assume that the
IGM is photoionized and photoheated by a UV background close to that
inferred from quasars (as computed by Haardt \& Madau 1996, hereafter
HM96). The overall normalization of the column density distribution
depends approximately on the parameter $\Omega_b^2 h^3/\Gamma$, where
$\Gamma$ is the hydrogen ionization rate, $h$ is the Hubble constant
in units of $100\,\kmsmpc$, and $\Omega_b$ is the baryon density
parameter. The width of the lines, as measured by the $b$ parameter of
a Voigt profile, is set by thermal broadening, peculiar velocities,
and Hubble expansion across the filaments.

While the first simulations showed good agreement with the observed
line statistics, recent detailed studies at higher numerical
resolution have revealed a serious conflict with the data: the models
predict median values for the $b$ distribution of about $20\,\kms$ at
$z=3$, compared with the observed median of $30\,\kms$ (Theuns \etal
1998b; Bryan \etal 1999). One possible way to restore the agreement
with observations is to increase the temperature of the IGM, adding to
the thermal broadening.  As Theuns \etal (1999) have pointed out, this
may be achieved by increasing the baryon density $\omega_b \equiv
\Omega_b h^2$, or by heating the gas for longer by increasing the age
of the universe (e.g. by lowering the total matter density
$\Omega_m$).  However, the gas temperature at late times is determined
(approximately) by the balance between adiabatic cooling and
photoelectric heating and varies as $T\propto (\Omega_b
h/\sqrt{\Omega_m}) ^{1/1.7}$ (Hui \& Gnedin 1997). Thus the needed
increase in temperature (a factor of 2) is difficult to accomplish
without conflicting with big bang nucleosynthesis for reasonable
values of $\Omega_m$ and $h$ (Theuns \etal 1999 and \S~3 of this
paper).  Delaying the reionization of helium may also help (Haehnelt
\& Steinmetz 1998), but this appears to boost the temperature by only
a small factor. Other mechanisms that could increase
the line--width include photoelectric dust heating (Nath, Sethi, \& Shchekinov
1999) and radiative transfer effects during the reionization of
helium (Miralda--Escud\'e \& Rees 1994; Abel \& Haehnelt 1999).

In this {\it Letter} we point out that a potentially important heating
term has been neglected in all previous numerical studies, namely
Compton heating of electrons by hard X--ray background photons. We
show that this effect will dominate over photoelectric heating of
hydrogen at redshifts $\gta 2$ in popular models for the redshift
evolution of the extragalactic background radiation and that it will
drive the IGM towards a temperature of about $1.5\times 10^4\,$ K at
$z>3$.

\section{Compton heating of the IGM}

In an IGM that interacts solely with a background of UV ionizing photons
(assumed to originate from quasars and/or star--forming galaxies),     
the dominant heat input is due to photoelectric heating of hydrogen and 
helium atoms. The energy input from hydrogen photoionization
is given by 
\begin{equation}
H_\nH=4\pi \, {\bar n}_{\rm H\,I}\int_{\nu_L}^\infty d\nu \, {J_\nu\over 
h_P\nu}\, \sigma_\nH \,(h_P\nu-h_P\nu_L), \label{photo}
\end{equation}   
where $\bar{n}_\nHI$ is the mean neutral hydrogen density, $J_\nu\propto
(\nu/\nu_L)^{-\alpha}$ the specific intensity of the UV radiation 
background, 
$\sigma_\nH=\sigma_L (\nu/\nu_L)^{-3}$ the photoionization 
cross--section, and $h_P\nu_L$ the energy of the Lyman edge. Equation 
(\ref{photo}) can be integrated to give  
\begin{equation}
H_\nH={\bar n}_\nHI\, \sigma_L\, c\, U_{\rm UV}\,{(\alpha-1)\over 
(\alpha+2)(\alpha+3)}\, \label{photo2}
\end{equation}   
where $U_{\rm UV}$ is the (proper) energy density of the metagalactic flux.
In a quasar--dominated background, photoheating by \HeII will 
be larger than $H_\nH$ by a factor 1.5--2. The main cooling 
processes are Compton 
cooling against the microwave background at high redshifts and adiabatic 
cooling at redshifts $\lta 2$.

Allow now the intergalactic gas to exchange energy with the XRB through Compton
heating at a rate 
\begin{equation}
H_C=4\pi \, {\bar n}_e\, {\sigma_T\over m_ec^2}\int_0^\infty 
d\nu \, J_\nu\, (h_P\nu-4kT), \label{compt}
\end{equation}   
where $\bar{n}_e$ is the mean electron density, $J_\nu$ the specific 
intensity of the XRB, and $\sigma_T$ the Thomson cross--section.
Energy is transferred from photons to electrons when $h_P\nu\gg 4kT$,
and in this limit equation (\ref{compt}) becomes
\begin{equation}
H_C=\sigma_T\, {\bar n}_e\, c {\langle h_P\nu\rangle\over m_ec^2}U_X,
\label{compt2}
\end{equation}   
where $\langle h_P\nu\rangle$ is the mean photon energy and $U_X$   
the energy density of the XRB. For photons above 100 keV relativistic 
effects become significant, 
and Compton scattering takes place in the Klein--Nishina regime. The 
relativistic generalization of equation (\ref{compt}) can be written as
(cf. Blumenthal 1974)
\begin{equation}
H_C=3\pi \, {\bar n}_e\, \sigma_T\, 
\int_0^\infty d\epsi \, J_\epsi\, {1\over \epsi^2}\left[{\epsi^2 -2\epsi-3
\over 2\epsi} \ln(1+2\epsi)-{10\epsi^4-51\epsi^3-93\epsi^2-51\epsi-9
\over 3(1+2\epsi)^3}\right]
\label{comptKN}
\end{equation}   
(for $\epsi\gg 4kT/m_ec^2$), where $\epsi$ is the photon energy in units
of $m_ec^2$. All the energy of the primary electrons
 will be deposited into the
 medium as heat
(through Coulomb collisions), since energy losses due to collisional excitation
and ionization of H and He are negligible in such a highly ionized IGM 
(e.g. Shull \& Van Steenberg 1985).

The XRB spectrum is characterized by a narrow peak in spectral power, $EJ_E$,
at 30 keV; a simple analytical fit to the observed energy flux in units 
of $\xrbunits$ yields 
\begin{equation}
J_E = 7.7 E_{\rm keV}^{-0.29}\, \exp\left(-{E_{\rm keV}\over40}\right)
\end{equation}
from 3 to 60 keV (Boldt 1987). At higher energies, the {\it HEAO} 1 A--4 data 
can be fitted with a power--law,  
\begin{equation}
J_E= 0.26 \left({E_{\rm keV}\over100}\right)^{-1.75}
\end{equation}
from 80 to 400 keV (Kinzer \etal 1997). The preliminary
diffuse $\gamma$--ray spectrum from 800 keV to 30 MeV measured with COMPTEL
(Kappadath \etal 1996) obeys another power--law, 
\begin{equation} 
J_E= 0.15\, (E_{\rm keV}/100)^{-1.38}\,\xrbunits
\end{equation}
(Kribs \& Rothstein 1997). The total energy density in the 0.003--30
MeV band is $U_X=6.3\times 10^{-5}\,\eden$. The mean, ``effective''
(as it includes a large correction due to Klein--Nishina effects)
photon energy is $\langle h_P\nu\rangle = 31.8\,$ keV. Note that,
because of the decline at high energies of the Klein--Nishina
cross--section, the (X$+\gamma$)--ray heating rate is numerically
equivalent to the rate one would derive by assuming pure Thomson
scattering of photons in the range 3--180 keV only. With the adopted
spectrum, about half of the total energy exchange rate at the present--epoch
is provided by
photons above 110 keV and about 14\% is contributed by photons above
1 MeV.

We can now estimate the relative importance of Compton versus 
photoelectric heating of hydrogen. The QSO contribution to the local 
metagalactic flux at the Lyman edge is $J_L\approx 10^{-23}\,$ $\uvunits$ 
(HM96). Assuming a 
spectral index of $\alpha=1.3$, from equations (\ref{photo}) and (\ref{compt})
we get  
\begin{equation}
{H_C\over H_\nH}\approx 6.8\times 10^{-7}{\bar n_e\over \bar n_\nHI} 
\end{equation}
at $z=0$. The equation of ionization equilibrium for an IGM with  
mean hydrogen density $\bar{n}_\nH=1.6\times 10^{-7}$ $(\omega_b/0.019)$ 
cm$^{-3}$ and temperature $T=2\times 10^3$ K gives $\bar n_e/\bar n_\nHI\sim 
1.5\times 10^{5}$, hence $H_C\sim 0.1\, H_\nH$. At the present--epoch, 
photoelectric heating is then the dominant contribution. At $z=2$, the UV 
background from optically--selected quasars increases to $J_L\approx 3\times 
10^{-22}\,$ $\uvunits$ (HM96), the gas temperature reaches $10^4\,$K or 
higher, and ionization equilibrium  now yields $\bar n_e/\bar n_\nHI\sim 
5\times 10^{5}$. If the bulk of the XRB is produced at high 
redshifts, $U_X$ increases as $(1+z)^4$, and the rate of gain in 
thermal energy per baryon due to Compton scattering goes up as $(1+z)^{4
+\beta}$, where $\beta\approx 1/3$ is less than unity because Klein--Nishina 
corrections become more important at high redshifts. The same exercise then 
yields $H_C\sim 1.3\, H_\nH$ at $z=2$. 

The relative strength of the two competing heating terms will obviously depend 
on the redshift evolution of ionizing and X--ray extragalactic radiation.
Two cosmological effects may tend to make Compton scattering of hard X--ray
photons the dominant source of heating at high redshifts in a highly ionized, 
low density IGM:  
\begin{itemize}
\item at fixed ionization state and energy density ratio $U_X/U_{\rm UV}$, 
the ratio $H_C/H_\nH$ increases with $\langle h_P\nu\rangle$ as 
$(1+z)^{\beta}$, where $\beta\approx 1/3$ for $z\le 3$; 
\item at fixed ratio between the quasar UV and X--ray 
volume emissivities, the radiation energy density $U_X$ increases with 
redshift 
relative to $U_{\rm UV}$, as the mean free path of photons at $912\,$\AA\ 
becomes very small, sources at higher 
redshifts are severely attenuated, and the UV background is largely ``local''
beyond a redshift of $z\gta 2$ (e.g. Madau 1992).
\end{itemize}
In the absence of any other cooling or heating mechanisms, Compton cooling  
off cosmic microwave background (CMB) photons and Compton heating by the XRB
will drive the thermal state of the IGM towards the Compton temperature 
\begin{equation}
T_C={\langle h_P\nu\rangle\over 4 k}\, {U_X\over U_{\rm CMB}},
\end{equation}
independently of the electron density. At $z=0$, $T_C\approx 2.2\times 10^4\,$
K. 

\section{Non--equilibrium models}

In this section we will calculate numerically the thermal history of a
uniform IGM of primordial composition within a cosmological
context. The code that we use includes the relevant cooling and
heating processes (as described in Theuns \etal 1998b, and references
therein) and follows the non--equilibrium evolution of hydrogen and
helium ionic species. The gas is allowed to interact with the CMB
through Compton cooling, with the XRB through Compton heating, and
with the time--dependent QSO ionizing background as computed by HM96.

It is widely believed that the diffuse XRB is due to the summed
emission from AGNs (see Fabian \& Barcons 1992 for a review). There is
insufficient information, however, to uniquely infer from the data the
evolution with cosmic time of the hard X--ray volume emissivity.
According to the findings of Boyle \etal (1994) from {\it ROSAT} data
(see also Page \etal 1996), the model that best fits the cosmological
evolution of soft X--ray selected AGNs in an Einstein--deSitter
universe is one in which the comoving volume emissivity increases as
$(1+z)^{3.15}$ for $z<1.6$ (in qualitative agreement with the evolution 
of the QSO optical luminosity function), and remains roughly constant at 
higher redshifts, up to $z\sim 4$ (Miyaji, Hasinger, \& Schmidt
1998). Assuming that these simple prescriptions also
describe the evolution of hard X--ray sources, the energy density of
the XRB then increases approximately as $U_X\propto (1+z)^4$ at low
redshifts, and as $(1+z)^{2.5}$ at $z>1.6$. To account for the
existence of a high redshift cut--off, we will parameterize the
evolution of the X--ray energy density as
\begin{equation}
U_X(z)=U_X(0) (1+z)^4\, \exp(-z^2/z_c^2). \label{UX}
\end{equation}
Figure 1 shows the ensuing Compton heating rate per baryon as a function of 
redshift for $z_c=5$, compared with the photoionization heating rate from 
optically--selected quasars. Figure 2 shows the thermal history of an IGM with 
$\omega_b =0.019$ (Burles and Tytler 1998) for two cosmological models: 
an Einstein--deSitter universe and a spatially flat universe dominated by a
cosmological constant with parameters close to those inferred from
Type Ia supernovae and cosmic microwave background anisotropies (see
Efstathiou \etal 1999, and references therein). In both cases, we
assume a photoionizing background as computed by HM96 and a Hubble
constant of $h=0.65$ (see Freedman \etal 1998). The dotted lines show
the temperature evolution if the X--ray background is ignored, as has
been assumed in all previous work. The solid lines include Compton
heating by the XRB according to the model of equation (\ref{UX}).
Including the X--ray background with a redshift cut--off parameter of
$z_c=5$ causes the temperature of the IGM to rise to $1.4 \times
10^4\,$K at $z=4$ and $1.1 \times 10^4\,$K at $z=3$. 
The IGM temperatures are even higher in the
$\Lambda$--dominated model. 
These 
results are extremely insensitive to the amplitude of the
photoionizing background and relatively weakly dependent on $\omega_b$ 
and $h$. They are, however, sensitive to the assumed redshift
evolution of
the X--ray background and to the hard X--ray spectrum at photon
energies above $50$ keV, which is poorly known but dominate the heating rate.

The results of Figure 2 and the numerical simulations of Theuns \etal
(1999) suggest that heating of the IGM by the hard XRB may well be
able to account for the $b$--parameter distribution of the \Lya
forest.  Furthermore, our results suggest that the observed
$b$--parameter distribution may provide a useful probe of the
evolution of X--ray diffuse radiation at high redshift. Compton
heating will also tend to flatten the equation of state of the
intergalactic gas, since the Compton heating per unit mass is
independent of density. However, for our model with $z_c=5$, this
effect is small, even for low density regions.


\section{Discussion}

We have shown that Compton heating of electrons by hard X--ray
background photons provides an important heating source of the
IGM. For plausible parameters, heating by the XRB will be comparable
to photoelectric heating at redshifts $\gta 2$. This effect
has been ignored in most numerical simulations of the \Lya forest, yet
it can have a significant impact on the temperature of the IGM, raising
it to temperatures as high as $\sim 1.5 \times 10^4\,$K at redshifts $z
\gta 3$.

The increase in IGM temperature caused by this effect may explain why
the \Lya $b$--parameter distributions derived from numerical simulations
(Theuns \etal 1998b; Bryan \etal 1999) are narrower than those
observed. Theuns \etal (1999) have previously suggested that the
temperature of the IGM can be raised by invoking a high baryon density
in a universe with low total matter density. However, the density
required, $\omega_b \gta 0.025$, is well above the value implied
by primordial nucleosynthesis and the measurements of the deuterium
abundances from quasar spectra ($\omega_b = 0.0193 \pm 0.0014$,
Burles and Tytler 1998).  Compton heating by hard X--ray photons may
help to resolve this discrepancy. The origin and main production epoch
of the XRB are still largely unsolved problems. The $b$--parameter
distribution of the \Lya lines  could provide important information on
the evolution of X--ray background radiation at high redshifts.

Finally, it is interesting to note that, as another consequence of 
Compton heating of the IGM, there may exist an 
X--ray proximity effect, in analogy with the usual UV proximity effect
(e.g. Bechtold 1994 and references therein). This could
result in a hotter IGM, and consequently broader \Lya lines,
in the immediate vicinity of a long--lived X--ray emitting quasar
(lifetime $\gta$ the Compton heating timescale). 
 
\acknowledgments We have benefited from useful discussions with
G. Bryan, A. Meiksin, J. Miralda--Escud\'e, M. Rees, and T. Theuns.  We
are especially grateful to Jordi Miralda--Escud\'e for pointing out the
importance of relativistic corrections at high redshift. Support for this 
work was provided by NASA through ATP grant NAG5--4236, and by NSF through 
grant PHY94--07194(PM). GPE thanks PPARC for the award of a Senior 
Fellowship.

\references

Abel, T., \& Haehnelt, M. G. 1999, submitted to ApJ (astro--ph/9903102)

Bechtold, J. 1994, ApJS, 91, 1

Blumenthal, G. R. 1974, ApJ, 188, 121 

Boldt, E. 1987, Phys. Rep., 146, 215

Boyle, B. J., Shanks, T., Georgantopoulos, I., Stewart, G. C., \& Griffiths,
R. E. 1994, MNRAS, 271, 639

Bryan, G. L., Machacek, M., Anninos, P., \& Norman, M. L.  1999, ApJ, in press
(astro--ph/9805340)

Burles, S., \& Tytler, D. 1998, ApJ, 499, 699

Cen, R., Miralda--Escud\'e, J., Ostriker, J. P., \& Rauch, M. 1994, ApJ, 437, L83

Dav\'e, R., Hernquist, L., Katz, N., \& Weinberg, D. 1999, ApJ, 511, 521

Efstathiou, G., Bridle, S. L., Lasenby, A. N., Hobson M. P., \& 
Ellis, R. S. 1999, MNRAS, in press (astro--ph/9812226)

Fabian, A. C., \& Barcons, X. 1992, ARA\&A, 30, 429

Freedman, J. B., Mould, J. R., Kennicutt, R. C., \& Madore, B. F. 1998, 
preprint (astro--ph/9801080)

Haardt, F., \&  Madau, P. 1996, ApJ, 461, 20 (HM96)

Haehnelt, M., \& Steinmetz, M. 1998, MNRAS, 298, L21

Hernquist, L., Katz, N., Weinberg, D., \& Miralda--Escud\'e, J. 1996, ApJ, 457, 
L51

Hui, L., \& Gnedin, N. 1997, MNRAS, 292, 27

Kappadath, S. C., \etal 1996, A\&AS, 120, 619

Kinzer, R. L., Jung, G. V., Gruber, D. E., Matteson, J. L., \& Peterson,
L. E. 1997, ApJ, 475, 361

Kribs, G. D., \& Rothstein, I. Z. 1997, Phys. Rev. D., 55, 4435

Madau, P. 1992, ApJ, 389, L1

Miralda--Escud\'e, J., Cen, R., Ostriker, J. P., \& Rauch, M. 1996, ApJ, 471, 582

Miralda--Escud\'e, J., \& Rees, M. J. 1994, MNRAS, 266, 343

Miyaji, T., Hasinger, G., \& Schmidt, M. 1998, preprint (astro--ph/9809398)

Nath, B. B., Sethi, S. K., \& Shchekinov, Y. 1999, MNRAS, 303, 1

Page, M. J., \etal 1996, MNRAS, 281, 579


Shull, J. M., \& Van Steenberg, M.E. 1985, ApJ, 298, 268

Theuns, T., Leonard, A., \& Efstathiou, G.  1998a, MNRAS, 297, L49

Theuns, T., Leonard, A., Efstathiou, G., Pearce, F. R., \& Thomas, P. A. 1998b, 
MNRAS, 301, 478

Theuns, T., Leonard, A., Schaye J., \& Efstathiou, G. 1999, MNRAS, in press
(astro--ph/9812141)

Zhang, Y., Anninos, P., \& Norman, M. L. 1995, ApJ, 453, L57

Zhang, Y., Meiksin, A., Anninos, P., \& Norman, M. L. 1998, ApJ, 495, 63

\begin{figure}
\vskip 5.0 truein

\includegraphics{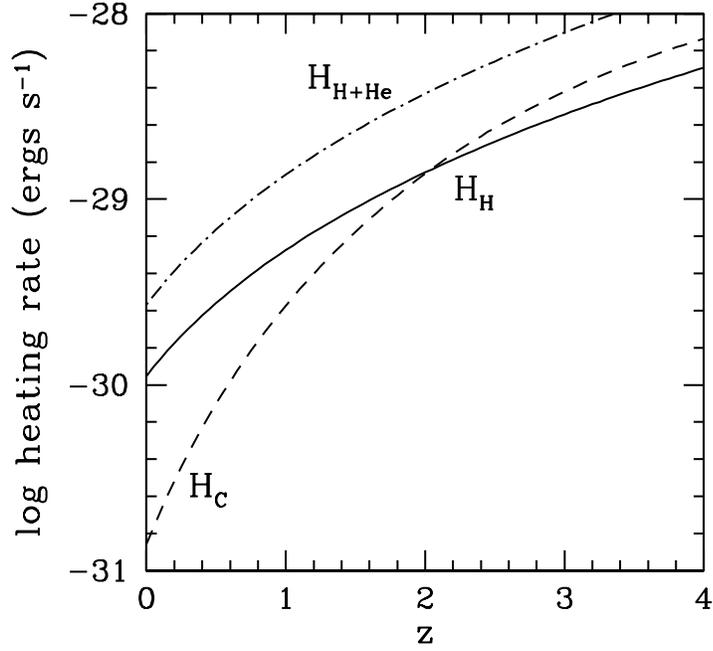}
\caption{\small Compton ({\it dashed line}) and photoionization ({\it
solid line:} H only; {\it dash--dotted line:} H$+$ He) heating rates
per baryon as a function of redshift. The rates have been derived assuming
ionization equilibrium in an IGM with $\omega_b=0.019$, the HM96 UV 
radiation field, the XRB evolution described in the text (with $z_c=5$),
and the corresponding temperature evolution shown in Figure 2{\it b}.
\label{fig1}}
\end{figure}
\vfill\eject

\begin{figure}

\vskip 5.0 truein

\includegraphics{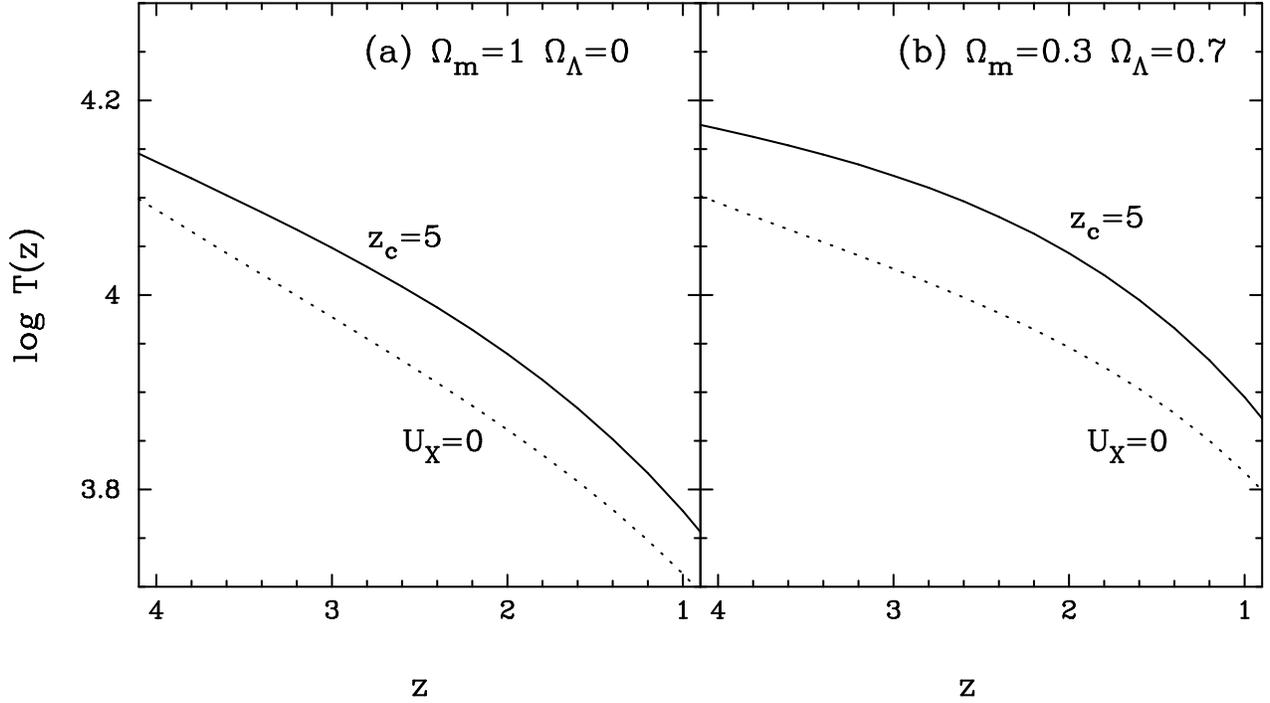}
\caption{\small Thermal history of a uniform intergalactic medium
for two spatially flat cosmologies. The dotted lines show the
temperature evolution when the X--ray background is ignored and the
only heating source is the quasar UV background as computed by HM96. 
The solid lines show the temperature evolution when the
XRB is included with redshift evolution as given by equation 
(\ref{UX}) for $z_c=5$. We have assumed $\omega_b=0.019$ and $h=0.65$.
\label{fig2}}
\end{figure}

%
%

\end{document}